\documentclass[%
 reprint,
superscriptaddress,
preprintnumbers,
 amsmath,amssymb,
 aps,prl,
]{revtex4-1}

\usepackage{float}
\restylefloat{table}

\usepackage{multirow}
\usepackage{color} 
\usepackage{graphicx}
\usepackage{dcolumn}
\usepackage{bm}

\usepackage{verbatim}
\usepackage{appendix}
\usepackage{amsthm}
\usepackage{hyperref}
\hypersetup{
colorlinks=true,
linkcolor=blue,
filecolor=blue,
citecolor=blue,  
urlcolor=black,
}

\theoremstyle{plain}
\newtheorem{thm}{Theorem}

\theoremstyle{definition}

\newcommand{\bra}[1]{{\langle#1|}}
\newcommand{\ket}[1]{{|#1\rangle}}

\newcommand{\ketbra}[2]{{\ket{#1}\!\bra{#2}}}
\newcommand{\norm}[1]{\left\lVert#1\right\rVert}

\newcommand{\tr}{\operatorname{tr}}
\DeclareMathOperator{\bbE}{\mathbb{E}}

\DeclareMathOperator{\Hom}{Hom}

\newcommand{\rmd}{\mathrm{d}}

\newcommand{\rmU}{\mathrm{U}}

\newcommand{\bbC}{\mathbb{C}}

\newcommand{\cat}{\mathrm{Cat}}

\begin{document}
\preprint{MIT-CTP/4935}
\title{Generalized Entanglement Entropies of Quantum Designs\\
}
\author{Zi-Wen Liu}\email{zwliu@mit.edu}
\affiliation{Center for Theoretical Physics, Massachusetts Institute of Technology, Cambridge, Massachusetts 02139, USA}
\affiliation{Department of Physics, Massachusetts Institute of Technology, Cambridge, Massachusetts 02139, USA}
\author{Seth Lloyd}
\affiliation{Department of Mechanical Engineering, Massachusetts Institute of Technology, Cambridge, Massachusetts 02139, USA}
\affiliation{Department of Physics, Massachusetts Institute of Technology, Cambridge, Massachusetts 02139, USA}
\author{Elton Yechao Zhu}
\affiliation{Center for Theoretical Physics, Massachusetts Institute of Technology, Cambridge, Massachusetts 02139, USA}
\affiliation{Department of Physics, Massachusetts Institute of Technology, Cambridge, Massachusetts 02139, USA}
\author{Huangjun Zhu}
\affiliation{Institute for Theoretical Physics, University of Cologne, 50937 Cologne, Germany}
\affiliation{Department of Physics and Center for Field Theory and Particle Physics, Fudan University, Shanghai 200433, China}
\affiliation{Institute for Nanoelectronic Devices and Quantum Computing, Fudan University, Shanghai 200433, China}
\affiliation{State Key Laboratory of Surface Physics, Fudan University, Shanghai 200433, China}
\affiliation{Collaborative Innovation Center of Advanced Microstructures, Nanjing 210093, China}

\date{\today}

\begin{abstract}
The entanglement properties of random quantum states or dynamics are important to the study of a broad spectrum of disciplines of physics, ranging from quantum information to high energy and many-body physics. This work investigates the interplay between the degrees of entanglement and randomness in pure states and unitary channels.
We reveal strong connections between designs (distributions of states or unitaries that match certain moments of the uniform Haar measure) and generalized entropies (entropic functions that depend on certain powers of the density operator), by showing that R\'enyi entanglement entropies averaged over designs of the same order are almost maximal. This strengthens the celebrated Page's theorem. 
Moreover, we find that designs of an order that is logarithmic in the dimension maximize all R\'enyi entanglement entropies, and so are completely random in terms of the entanglement spectrum. Our results relate the behaviors of R\'enyi entanglement entropies to the complexity of scrambling and quantum chaos in terms of the degree of randomness, and suggest a generalization of the fast scrambling conjecture.
\end{abstract}

\maketitle


\emph{Introduction.} The interplay between entanglement and randomness plays important roles in many areas of physics. A particular notion of wide interest is ``scrambling,'' which describes the phenomenon that initially localized quantum information spreads over the entire system via global entanglement, so that the information is lost from the perspective of any local observer, or the state of the system is effectively randomized.  
The concept of scrambling originates from the study of black holes and quantum gravity \cite{page93,mirror,fast,add}, and similar mechanisms also underlie many other key concepts in physics, such as quantum chaos \cite{PhysRevLett.80.5524,PhysRevE.64.036207,chaos}, quantum thermalization \cite{mblrev,popescu}, quantum data hiding \cite{PhysRevLett.86.5807,dh02}.
The entanglement properties of random or pseudorandom quantum states and channels can illuminate such phenomena, and are fundamental to relevant studies.

It has long been noted that a random state is typically highly entangled \cite{lubkin,lloydpagels}. This observation is formalized by the Page's theorem \cite{page93,FoonK94, Sanc95, Sen96}, which states that the expected von Neumann entropy of small subsystems of a completely random state (drawn from the Haar measure) is very close to the maximum.  Similar observations for the entanglement in random unitary channels are made recently in \cite{chaos}.
However, such results are not ``tight'' from the perspective of complexity. 
On the one hand, the complexity of Haar randomness is high: the number of local gates needed to even approximate the Haar distribution grows exponentially in the number of qubits \cite{1995quant.ph..8006K}. On the other hand, however, pseudorandom distributions with low complexity \cite{brandaoharrow1,brandaoharrow2,PhysRevA.75.042311,nakata} are sufficient to acquire the Page-like entanglement property.  
That is, there is a significant complexity gap between complete randomness and large entanglement entropy.  In dynamical scenarios, this gap corresponds to a substantial but poorly understood regime beyond scrambling, chaos and thermalization, where the randomness and complexity of the system can keep growing. Indeed, the common characteristics of information scrambling, such as global entanglement \cite{mirror,chaos}, remote signaling \cite{fast}, local indistinguishability \cite{Lashkari2013}, do not need nor imply complete randomization, and there is little knowledge about the physics of later times. 

To fill this gap, we consider more stringent entanglement measures. The study is also extended to unitary channels via the Choi isomorphism.  More specifically, we employ various techniques from representation theory, random matrix theory, combinatorics and Weingarten calculus to analytically study the generalized entanglement entropies (which depend on higher powers of the reduced density operator) of random and pseudorandom states and unitaries. 
A key collective finding is that the R\'enyi-$\alpha$ entanglement entropy averaged over $\alpha$-designs is almost maximal, where $\alpha$-designs stand for evenly distributed ensembles of states or unitaries that mimic the first $\alpha$ moments of the Haar measure, in analogy to $\alpha$-wise independent distributions which have wide applications in classical computer science and combinatorics.  In other words, designs represent finite-degree approximations of the truly random distribution on states or unitaries, which are of great interest in quantum information.  
This result links the order of entanglement entropies and that of designs, and closes the complexity gap in Page-like theorems. It also suggests R\'enyi entanglement entropies as diagnostics of the randomness complexity of corresponding designs in scrambling, as well as (truly quantum) witnesses of quantum pseudorandomness. 
The infinite order limit of R\'enyi entropy, which only depends on the largest eigenvalue, is known as the min entropy. 
We further show that the min entanglement entropy (and therefore all R\'enyi entanglement entropies) becomes almost maximal, which we also call ``max-scrambling'', for designs of an order that is only logarithmic in the dimension of the system. So designs of higher orders are essentially completely random in terms of entanglement. 
This leads to a strong estimate of the time needed to achieve max-scrambling based on the fast scrambling \cite{fast} and design \cite{nakata} conjectures.
Finally, we are able to construct state 2-designs such that all R\'enyi entanglement entropies of orders greater than 2 are bounded away from the maximum, which establishes an explicit separation between the complexities diagnosed by R\'enyi entanglement entropies.

This letter distills the key quantum information results of \cite{Liu2018}, which is written from the perspective of scrambling complexity.
Please refer to \cite{Liu2018} for technical details and more discussions.

\emph{Preliminaries.} Here we recall the formal definitions of designs and generalized entropies, the central mathematical concepts of this study.

Designs are ensembles of quantum states (unitaries) that are evenly distributed on the complex unit sphere (unitary group). 
They are efficient to implement \cite{brandaoharrow1,brandaoharrow2,PhysRevA.75.042311,nakata} and useful in many important quantum information processing tasks such as randomized benchmarking \cite{PhysRevLett.106.180504,1367-2630-16-10-103032} and decoupling \cite{1367-2630-15-5-053022}. There are several ways to characterize exact or approximate designs (see e.g.~\cite{rep}), among which the one based on polynomials is the most relevant to this work.
Let $\Hom_{(t,t)}(\bbC^d)$ be the space of polynomials homogeneous of degree $t$ both in the coordinates of vectors in $\bbC^d$ and in their complex conjugates. An ensemble $\nu$ of pure state vectors in dimension $d$ is a (complex projective) \emph{$t$-design} if 

  	\begin{equation*}
\bbE_\nu p(\psi)= \int \rmd \psi p(\psi)\quad \forall p\in \Hom_{(t,t)}(\bbC^d),
  	\end{equation*}
  	where $\bbE_\nu$ denotes the expectation value over $\nu$. The integral is taken with respect to the (normalized) uniform  measure on the complex unit sphere in $\bbC^d$. 
Designs of unitary channels can be defined analogously.  Let $\Hom_{(t,t)}(\rmU(d))$ be the space of polynomials homogeneous of degree $t$ both in the matrix elements of $U\in \rmU(d)$
  and in their complex conjugates. 
An ensemble $\mu$ of  unitary operators  in dimension $d$  is a  \emph{unitary $t$-design} if 
\begin{equation*}\label{eq:U2design}
\bbE_\mu p(U) =\int \rmd U p(U) \quad \forall p\in \Hom_{(t,t)}(\rmU(d)),
\end{equation*}
where the integral is taken over the normalized Haar measure on ${\rm U}(d)$.

 Order-$\alpha$ entropies of a density operator $\rho$ are entropic functions (which we call characteristic functions) of $\mathrm{tr}\{\rho^\alpha\}$. A unified definition of such entropies is given by
$S^{(\alpha)}_s(\rho)=\frac{1}{s(1-\alpha)}\left[(\mathrm{tr}\{\rho^\alpha\})^s-1\right]$,
where $s$ is a parameter that identifies the characteristic function and the family of entropies.
The most representative families are R\'enyi (the limiting case $s\rightarrow 0$) and Tsallis ($s=1$) entropies. In this work, we mostly focus on R\'enyi entropies
\begin{equation*}
S^{(\alpha)}_R(\rho)=\frac{1}{1-\alpha}\log\mathrm{tr}\{\rho^\alpha\},
\end{equation*}
with  orders $\alpha$ being  positive integers.
In contrast to other generalized entropies, R\'enyi entropies have the following desirable properties, which make this family most relevant. First, they are convex in $\mathrm{tr}\{\rho^\alpha\}$, which makes it possible to use Jensen's inequality to lower bound the design-averaged values by Haar integrals. Second, they have the same roof value $n$ for uniform spectrum for systems of $n$ qubits, which allows meaningful comparisons with the maximum and between different orders. Third, they are additive on product states (otherwise it is not natural to define generalized quantities such as mutual information and tripartite information).
Several other properties of R\'enyi entropies also constitute the basis for our idea. First, when the order $\alpha$ increases, $S^{(\alpha)}_R$ becomes more and more sensitive to the nonuniformity in the spectrum: $S^{(\alpha_1)}_R\geq S^{(\alpha_2)}_R$ when $\alpha_1<\alpha_2$.
In particular, taking the $\alpha\rightarrow\infty$ limit yields the min entropy:
$$  S_{\min}(\rho)=-\log\norm{\rho} =-\log\lambda_{\max}(\rho), $$
where $\norm{\cdot}$ denotes the operator norm and $\lambda_{\max}(\cdot)$ denotes the largest eigenvalue. Min entropy lower bounds all R\'enyi entropies.  Second, for any positive integer $\alpha$, there exist distributions such that the R\'enyi-$\alpha$ entropies are very close to the maximum (the gap is $O(1)$), but the R\'enyi entropies of all higher orders are bounded away from the maximum (the gap grows as $n$).
This ``cutoff'' phenomenon allows the possibility of separating the complexities of scrambling by R\'enyi entropies. 
Third, the gap between the R\'enyi entropies and the maximum cannot increase under partial trace.  So the near-maximality of the R\'enyi entanglement entropies of half-half partitions ensures that those of all partitions are almost maximal.
See \cite{Liu2018} for details of the above arguments.

\emph{Random states.} We first introduce results on random pure states. 
Consider a bipartite system with Hilbert space $\mathcal{H}=\mathcal{H}_A\otimes\mathcal{H}_B$, where $\mathcal{H}_A$ and $\mathcal{H}_B$ have dimensions $d_A$ and $d_B$, respectively. 
The entanglement entropy between partitions $A$ and $B$ of a pure state $\ket{\psi}$ is given by the entropy of the reduced density operator $\rho_{A} = \mathrm{tr}_B(\ketbra{\psi}{\psi})$. 

A key observation is that, given an $\alpha$-design $\nu_\alpha$, we have
$ \bbE_{\nu_\alpha}\mathrm{tr}\{\rho_{A}^\alpha\} = \int \rmd\psi\mathrm{tr}\{\rho_{A}^\alpha\}$
since $\mathrm{tr}\{\rho_{A}^\alpha\}$ only involves $\Hom_{(\alpha,\alpha)}$ terms of the entries of $\ket{\psi}$. Since the characteristic function for the R\'enyi-$\alpha$ entropy is convex, $\bbE_{\nu_\alpha} S^{(\alpha)}_R(\rho_A)$ is lower bounded by the characteristic function of the Haar integral $\int \rmd\psi\mathrm{tr}\{\rho_{A}^\alpha\}$ by Jensen's inequality.
Calculation shows that 
\begin{equation}\label{eq:statetr}
    \int \rmd\psi\mathrm{tr}\{\rho_{A}^\alpha\} = \frac{1}{\alpha!D_{[\alpha]}}\sum_{\sigma\in S_\alpha}d_A^{\xi(\sigma\tau)}d_B^{\xi(\sigma)},
\end{equation}
where 
$
D_{[\alpha]}=\binom{d_Ad_B+\alpha-1}{\alpha}$
is the dimension of the symmetric subspace of $\mathcal{H}^{\otimes \alpha}$, $S_\alpha$ is the symmetric group of $\alpha$ symbols, $\xi(\sigma)$ is the number of disjoint cycles associated with $\sigma$ \footnote{Every element of the symmetric group can be uniquely decomposed into a product of disjoint cycles (up to relabeling).}, and $\tau:=(1~2~\cdots ~\alpha)$ is the 1-shift (canonical full cycle). 
We noticed that similar results have been derived and rederived several times \cite{lubkin,ZyczS01,MalaML02, Collins2010,CollN11}. A simple derivation was presented in \cite{Liu2018}.

First, consider equal partitions $d_A=d_B$ and the limit of large dimension. Here we introduce the following cycle lemma (proof in \cite{Liu2018},  cf.~\cite{nica2006lectures}): for all $\sigma\in S_\alpha$, $\xi(\sigma\tau)+\xi(\sigma)\leq\alpha+1$. Then Eq.~(\ref{eq:statetr}) reduces to
\begin{equation}
\int \rmd\psi\mathrm{tr}\{\rho_{A}^\alpha\}=\cat_\alpha d_A^{-\alpha+1}+O\Bigl(d_A^{-(\alpha+1)}\Bigr),
\end{equation}
where $\cat_\alpha$ is the $\alpha$-th Catalan number, satisfying $\frac{\log \cat_\alpha}{\alpha-1} \leq 2$ for all $\alpha\geq 2$.
So we obtain the following Theorem:
\begin{thm}
Let $\nu_\alpha$ be a projective $\alpha$-design. Consider equal partitions $d_A=d_B$. 
As $d_A\rightarrow\infty$,
\begin{equation}
\bbE_{\nu_\alpha} S^{(\alpha)}_R(\rho_A)\geq\log d_A-\frac{\log \cat_\alpha}{\alpha-1} +O(d_A^{-2}).
\end{equation}
So,
\begin{equation}
\bbE_{\nu_\alpha} S^{(\alpha)}_R(\rho_A)\geq\log d_A - O(1).
\end{equation}
\end{thm}
\noindent That is, the R\'enyi-$\alpha$ entanglement entropy across any cut averaged over an $\alpha$-design is very close to (at most a constant away from) the maximum. 

In fact, we are able to derive explicit bounds for finite dimensions and non-equal partitions:
\begin{thm}
Let $\nu_\alpha$ be a projective $\alpha$-design.
Let $q:=\alpha^3/(32d_B^2)<1, h(q):=1+2q/[3(1-q)]$.
For any $d_A\leq d_B$ and all $\alpha$,
\begin{eqnarray}
	\bbE_{\nu_\alpha} S^{(\alpha)}_R(\rho_A)&\geq& \log d_A-\frac{2\alpha-\frac{3}{2}\log\alpha +\log h(q) -\frac{1}{2}\log\pi  }{\alpha-1}\nonumber\\ &\geq& \log d_A-2. 
	\end{eqnarray}
	We also obtain the following bound, which improves the above result when $d_A$ is small:
	\begin{eqnarray}
	\bbE_{\nu_\alpha} S^{(\alpha)}_R(\rho_A)&\geq& \log d_A-2\log\left(1+\sqrt{\frac{d_A}{d_B}}\right)-\log c\nonumber\\&\geq&\log
	d_A-\frac{2}{\ln 2}\sqrt{\frac{d_A}{d_B}}-\log c,
	\end{eqnarray}
	where $c=1$ if  $\mathcal{H}$ is real and   $c=2$ if $\mathcal{H}$ is complex.

\end{thm}
 
 Error bounds indicating that the above results are highly robust against small deviations from exact designs can be found in \cite{Liu2018}.
 These results can be regarded as improved Page's theorems that are tight in terms of the complexity.

 Now we focus on the min entropy, given by $\alpha\rightarrow\infty$. Large min entropy implies that the spectrum is almost completely uniform. Are designs of infinite orders needed to achieve almost maximal min entanglement entropy?  The following result answers the question in the negative: 
 \begin{thm}
 Let $\nu_\alpha$ be a projective $\alpha$-design, where  $\alpha=\lceil (\log d_A)/a\rceil\leq (16d_B^2)^{1/3}$ with $0<a\leq 1$. 
	Then
\begin{equation}
\bbE_{\nu_\alpha} S_{\min}(\rho_A)\geq \log d_A-2-a.
\end{equation}
In particular,  $\bbE_{\nu_\alpha} S_{\min}(\rho_A)\geq \log d_A-3$ if $\alpha=\lceil \log d_A\rceil$. 
 \end{thm}
 \noindent That is, $\Omega(\log d_A)$-designs maximize all R\'enyi entanglement entropies, and so are essentially indistinguishable from the Haar measure by the entanglement spectrum.
 
Conversely, one may wonder whether there exist $\alpha$-designs such that R\'enyi entanglement entropies of orders larger than $\alpha$ are bounded away from the maximum, which we call ``gap $\alpha$-designs''. This indicates that they do not behave like designs of higher orders in a strong sense. 
 Here we present an explicit example of gap 2-designs. 
Let $G=\rmU_A\otimes \rmU_B$, where $\rmU_A, \rmU_B$ are the unitary groups on $\mathcal{H}_A, \mathcal{H}_B$, respectively. Calculation shows that the orbit of 
$|\psi\rangle $ under the action of $G$ forms a 2-design if and only if  $\tr\{\rho_A^2\}$ with $\rho_A=\tr_B(|\psi\rangle\langle\psi|)$ is  equal to the average over the uniform ensemble, that is,
\begin{equation}\label{eq:2-designCon}
\tr\{\rho_A^2\}=\frac{d_A+d_B}{d_Ad_B+1}.
\end{equation}
The same conclusion still holds if $\rmU_A, \rmU_B$
are replaced by subgroups that form unitary 2-designs on $\mathcal{H}_A, \mathcal{H}_B$, respectively. Equation~\eqref{eq:2-designCon} holds if $\rho_A$ has the following spectrum
\begin{align*}
\lambda_1=\frac{d_Ad_B+1+(d_A-1)\sqrt{(d_A+1)(d_Ad_B+1)}}{d_A(d_Ad_B+1)},\\ \lambda_2=\cdots=\lambda_{d_A} =\frac{d_Ad_B+1-\sqrt{(d_A+1)(d_Ad_B+1)}}{d_A(d_Ad_B+1)}.
\end{align*}
Suppose $d_B/d_A\leq r$ where $r$ is a constant, then $\lambda_1\geq  (rd_A)^{-1/2}$, and so
\begin{equation}
S_R^{(\alpha)}(\rho_A)\leq \frac{1}{1-\alpha}\log\lambda_1^\alpha\leq\frac{\alpha}{2(\alpha-1)}(\log d_A +\log r).
\end{equation}
As $d_A$ increases, the gap of $S_R^{(\alpha)}(\rho_A)$ from the maximum is unbounded for all $\alpha>2$.

\emph{Random unitary channels.} Now we extend the above analysis of pure states to the intrinsic entanglement properties of random unitary channels.  The key results are similar in spirit to those for states, although the derivations are  considerably more involved.

The Choi isomorphism (more generally, the channel-state duality) is widely used in quantum information theory to study quantum channels as states, by which a unitary operator $U$ acting on a $d$-dimensional Hilbert space $U=\sum_{i,j=0}^{d-1}U_{ij}\ket{i}\bra{j}$ is dual to the pure state
\begin{equation*}\label{choi}
    \ket{U} = \frac{1}{\sqrt{d}}\sum_{i,j=0}^{d-1}U_{ji}\ket{i}_{\mathrm{in}}\otimes\ket{j}_{\mathrm{out}},
\end{equation*}
which is called the Choi state of $U$. 
Consider bipartitions of the input register into $A$ and $B$, and the output register into $C$ and $D$. Let $d_A,d_B,d_C,d_D$ be the dimensions of subregions $A,B,C,D$, respectively ($d_A d_B= d_C d_D = d$).
We study the entropy of $\rho_{AC}$ with $\rho_{AC} = \mathrm{tr}_{BD}(\ket{U}\bra{U})$. 
 Consider the negative tripartite information
\begin{equation*}
    -I_3(A:C:D) := I(A:CD) - I(A:C) - I(A:D),
\end{equation*}
which is suggested in \cite{chaos} to diagnose information scrambling, since it intuitively measures the delocalization of local information.
Here $I(A:C)=S(A)+S(C)-S(AC)$ is the mutual information, which measures the total correlation between $A$ and $C$. Since the input and output are maximally mixed due to unitarity, the four subregions are all maximally mixed. As a result, $-I_3$ is determined by the entanglement entropy $S(AC)$. Indeed, $-I_3$ essentially measures the ability of a channel to generate global entanglement that ``hides'' the delocalized information. 
 Note that $-I_3$ can be reduced to the conditional mutual information $I(A:B|C)$ \cite{ding}, which is of great interest in quantum information theory. 


Given a unitary $\alpha$-design $\mu_\alpha$. By similar arguments involving the definition of unitary designs and the convexity of the R\'enyi characteristic function, the problem of bounding $\bbE_{\mu_\alpha} S_R^{(\alpha)}(\rho_{AC})$ boils down to computing the Haar integral $\int\rmd U\mathrm{tr}\{\rho_{AC}^\alpha\}$. In general, we find that
\begin{eqnarray}\label{eq:tru}
   && \int{\rm d}U \mathrm{tr}\left\{\rho_{AC}^\alpha\right\}\nonumber\\
&=&\frac{1}{d^{\alpha}}\sum_{\sigma,\gamma\in S_\alpha}d_A^{\xi(\sigma\tau)}d_B^{\xi(\sigma)}d_C^{\xi(\gamma\tau)}d_D^{\xi(\gamma)}\mathrm{Wg}(d,\sigma\gamma^{-1}),\label{general}
\end{eqnarray}
where
\begin{equation*}
\mathrm{Wg}(d,\sigma) = \frac{1}{(\alpha!)^2}\sum_{\lambda\vdash\alpha}\frac{\chi^\lambda(1)^2\chi^\lambda(\sigma)}{s_{\lambda,d}(1,\cdots,1)}
\end{equation*}
are Weingarten functions of $\mathrm{U}(d)$. Here $\lambda\vdash\alpha$ means $\lambda$ is a partition of $\alpha$, $\chi^\lambda$ is the corresponding character of $S_\alpha$, and $s_\lambda$ is the corresponding Schur function/polynomial. Notice that $s_{\lambda,d}(1,\cdots,1)$ is simply the dimension of the irrep of $\mathrm{U}(d)$ associated with $\lambda$.
The Weingarten function can be derived by various tools in representation theory, such as Schur-Weyl duality \cite{collins,collins2} and Jucys-Murphy elements \cite{zinn}.

For equal partitions, in the limit of large dimension, we obtain the following analogous result by applying the cycle lemma:
\begin{thm}
 Let $\mu_\alpha$ be a unitary $\alpha$-design.
 Consider equal partitions of the input and output registers, $d_A=d_B=d_C=d_D$. As $d\rightarrow\infty$,
\begin{equation}
\bbE_{\mu_\alpha} S_R^{(\alpha)}(\rho_{AC})\geq \log d - \frac{\log \cat_\alpha}{\alpha-1} + O(d^{-1}).
\end{equation}
So,
\begin{equation}
    \bbE_{\mu_\alpha} S_R^{(\alpha)}(\rho_{AC})\geq \log d - O(1).
\end{equation}
\end{thm}
\noindent Therefore, the R\'enyi-$\alpha$ entanglement entropy between $AC$ and $BD$ (and the corresponding negative tripartite information based on the R\'enyi-$\alpha$ entropy) averaged over unitary $\alpha$-designs is almost maximal.

We also provide explicit bounds for finite dimensions:
\begin{thm}
 Let $\mu_\alpha$ be a unitary $\alpha$-design.	 Suppose $d>\sqrt{6}\alpha^{7/4}$ and $d_A\leq d_B$. Then
	\begin{eqnarray}
&&\bbE_{\mu_\alpha} S_R^{(\alpha)}(\rho_{AC})\nonumber\\&\geq& \log d-\frac{\log \cat_\alpha}{\alpha-1}-\frac{\log \left[\frac{a_\alpha h(q) }{8}\left(7+\cosh\frac{2\alpha(\alpha-1)}{d}\right)\right]}{\alpha-1},\nonumber\\ \label{finited2}
	\end{eqnarray}
where $a_\alpha:=\left(1-\frac{6\alpha^{7/2}}{d^2}\right)^{-1}$.
\end{thm}

Similarly, these results do not deviate much for approximate unitary designs (see \cite{Liu2018} for detailed error analysis).


The result on the min entropy is also similar:
\begin{thm}
 Let  $\mu_\alpha$ be a unitary $\alpha$-design, where	$1\leq\alpha= \lceil \log d/a\rceil \leq \sqrt{d}/2$ and $ a> 0$; then 
	\begin{equation}
	\bbE_{\nu_\alpha} S_{\min}(\rho_{AC})\geq \log d-2-a.  \label{eq:AveMinEntropyChoiLog2}   
	\end{equation}	
In particular, $\bbE_{\nu_\alpha} S_{\min}(\rho_{AC})\geq \log d-3$ if $\alpha\geq \lceil \log d\rceil$. 
 \end{thm}
 \noindent Therefore,  unitary $\Omega(\log d)$-designs maximize all R\'enyi entanglement entropies.

 \emph{Design complexities by R\'enyi.} In the above we presented kinematic results revealing fundamental correspondences between R\'enyi entanglement entropies and quantum designs, which imply that states or unitaries sampled from $\alpha$-designs typically exhibit nearly maximal R\'enyi-$\alpha$ entanglement entropy. This also suggests R\'enyi-$\alpha$ entanglement entropy as potential diagnostics of the randomness complexity of $\alpha$-designs beyond information scrambling, in dynamical scenarios. Note that a recent work \cite{Roberts2017} generalizes the out-of-time-order correlators (which are widely used in the study of scrambling, see e.g.~\cite{Shenker2014,Shenker20142,Shenker2015,Roberts2015,PhysRevLett.115.131603,Maldacena2016,weakchaos}) to higher points, and establish similar connections to unitary designs via frame potentials.
 
 In particular, the saturation of the min entanglement entropy indicates that the system looks completely random (and the local information is completely lost) to any local observer, which is the strongest form of scrambling that we call ``max-scrambling''. How fast can physical systems achieve max-scrambling? The recent design Hamiltonian conjecture \cite{nakata} argues, based on the original fast scrambling conjecture \cite{fast}, that there exist physical dynamics (represented by local \footnote{Here ``local'' should mean all interaction terms involve at most $k$ qubits where $k$ is some constant ($k$-local), rather than spatially local on a finite-dimensional graph. This was not made very clear in \cite{nakata}.} and time-independent random Hamiltonians) that achieve approximate unitary $\alpha$-designs in $O(\alpha\log n)$ time, where $n$ is the number of qubits.
 Our result on logarithmic designs indicates that $O(n)$-designs are sufficient for max-scrambling, and therefore suggest the following fast max-scrambling conjecture: max-scrambling can be achieved by physical dynamics in ${O}(n\log n)$ time.  
 
 \emph{Outlook.} The mathematical results of this letter concern the average R\'enyi entanglement entropies of state and unitary designs. Some technical problems are left open. For example, we are not yet able to construct gap $\alpha$-designs for $\alpha>2$ and for unitaries. Moreover, due to the lack of subadditivity, the negative tripartite information $-I_3$ in terms of R\'enyi entropies is not necessarily positive. It is worth looking into when this situation occurs, and further considering the meanings of such derived quantities.  Also, the results here are about expected values. It would be nice to further analyze the variances or derive probabilistic bounds on concentration to talk about ``typical'' behaviors in a more rigorous sense.
 
 Our results suggest R\'enyi entanglement entropies as powerful tools to further advance the study of quantum randomness and pseudorandomness.
 For example, a particularly interesting insight is that R\'enyi entropies of non-integer orders are naturally defined, which indicates that they can be helpful in understanding the mysterious but potentially important notion of non-integer designs. 
 The techniques and results may find more applications in relevant areas in quantum information, such as entanglement theory, quantum complexity theory, quantum computing, and quantum cryptography.

 The physical aspects are certainly worth further exploration. For example, it would be interesting to study the dynamical behaviors of R\'enyi entanglement entropies and randomness in specific many-body or holographic systems, to learn about the physics in the post-scrambling regime and extend existing studies of entanglement growth (e.g., ``entanglement tsunami'' \cite{tsunami,PhysRevD.89.066012}). 
 A recent study \cite{Gu2017} on (a 1d variant of) the strongly chaotic SYK model \cite{PhysRevLett.70.3339,k} (which has drawn considerable interest as a solvable toy model of quantum black holes and holography) shows that, after a quench, there is a ``prethermal'' regime where light modes rapidly scramble, but the R\'enyi entanglement entropies do not reach thermal values, which confirms our expectation that the randomness complexity of the system is still low; however the late-time behaviors remain open. 
Moreover, it would be nice to extend the techniques and results of this work to the finite temperature regime or systems with conserved quantities, so as to apply our ideas in such physical scenarios and in general the study of quantum thermalization and many-body localization more carefully.  
We also hope to establish more solid connections between the randomness complexities and the conventional ones, such as computational, gate and Kolmogorov complexities, which play active roles in recent studies of holographic duality and black holes \cite{cealong,CEA,secondlaw}, and are of independent interest.
In general, the study of randomness complexities may also shed new light on the fruitful idea of modeling complex systems (especially black holes \cite{mirror}) by random states or dynamics.
 Further research along these lines are essential to our understanding of quantum chaos, quantum statistical mechanics, quantum many-body physics, and quantum gravity.


\begin{acknowledgements}
ZWL and SL are supported by AFOSR and ARO. EYZ is supported by the National Science Foundation under grant Contract Number CCF-1525130. Research at MIT CTP is supported by DOE. HZ is supported by the Excellence Initiative of the German Federal and State Governments (ZUK~81) and the DFG. 
\end{acknowledgements}


\begin{thebibliography}{53}%
\makeatletter
\providecommand \@ifxundefined [1]{%
 \@ifx{#1\undefined}
}%
\providecommand \@ifnum [1]{%
 \ifnum #1\expandafter \@firstoftwo
 \else \expandafter \@secondoftwo
 \fi
}%
\providecommand \@ifx [1]{%
 \ifx #1\expandafter \@firstoftwo
 \else \expandafter \@secondoftwo
 \fi
}%
\providecommand \natexlab [1]{#1}%
\providecommand \enquote  [1]{``#1''}%
\providecommand \bibnamefont  [1]{#1}%
\providecommand \bibfnamefont [1]{#1}%
\providecommand \citenamefont [1]{#1}%
\providecommand \href@noop [0]{\@secondoftwo}%
\providecommand \href [0]{\begingroup \@sanitize@url \@href}%
\providecommand \@href[1]{\@@startlink{#1}\@@href}%
\providecommand \@@href[1]{\endgroup#1\@@endlink}%
\providecommand \@sanitize@url [0]{\catcode `\\12\catcode `\$12\catcode
  `\&12\catcode `\#12\catcode `\^12\catcode `\_12\catcode `\%12\relax}%
\providecommand \@@startlink[1]{}%
\providecommand \@@endlink[0]{}%
\providecommand \url  [0]{\begingroup\@sanitize@url \@url }%
\providecommand \@url [1]{\endgroup\@href {#1}{\urlprefix }}%
\providecommand \urlprefix  [0]{URL }%
\providecommand \Eprint [0]{\href }%
\providecommand \doibase [0]{http://dx.doi.org/}%
\providecommand \selectlanguage [0]{\@gobble}%
\providecommand \bibinfo  [0]{\@secondoftwo}%
\providecommand \bibfield  [0]{\@secondoftwo}%
\providecommand \translation [1]{[#1]}%
\providecommand \BibitemOpen [0]{}%
\providecommand \bibitemStop [0]{}%
\providecommand \bibitemNoStop [0]{.\EOS\space}%
\providecommand \EOS [0]{\spacefactor3000\relax}%
\providecommand \BibitemShut  [1]{\csname bibitem#1\endcsname}%
\let\auto@bib@innerbib\@empty
\bibitem [{\citenamefont {Page}(1993)}]{page93}%
  \BibitemOpen
  \bibfield  {author} {\bibinfo {author} {\bibfnamefont {D.~N.}\ \bibnamefont
  {Page}},\ }\href {\doibase 10.1103/PhysRevLett.71.1291} {\bibfield  {journal}
  {\bibinfo  {journal} {Phys. Rev. Lett.}\ }\textbf {\bibinfo {volume} {71}},\
  \bibinfo {pages} {1291} (\bibinfo {year} {1993})}\BibitemShut {NoStop}%
\bibitem [{\citenamefont {Hayden}\ and\ \citenamefont
  {Preskill}(2007)}]{mirror}%
  \BibitemOpen
  \bibfield  {author} {\bibinfo {author} {\bibfnamefont {P.}~\bibnamefont
  {Hayden}}\ and\ \bibinfo {author} {\bibfnamefont {J.}~\bibnamefont
  {Preskill}},\ }\href {http://stacks.iop.org/1126-6708/2007/i=09/a=120}
  {\bibfield  {journal} {\bibinfo  {journal} {Journal of High Energy Physics}\
  }\textbf {\bibinfo {volume} {2007}},\ \bibinfo {pages} {120} (\bibinfo {year}
  {2007})}\BibitemShut {NoStop}%
\bibitem [{\citenamefont {{Sekino}}\ and\ \citenamefont
  {{Susskind}}(2008)}]{fast}%
  \BibitemOpen
  \bibfield  {author} {\bibinfo {author} {\bibfnamefont {Y.}~\bibnamefont
  {{Sekino}}}\ and\ \bibinfo {author} {\bibfnamefont {L.}~\bibnamefont
  {{Susskind}}},\ }\href {\doibase 10.1088/1126-6708/2008/10/065} {\bibfield
  {journal} {\bibinfo  {journal} {Journal of High Energy Physics}\ }\textbf
  {\bibinfo {volume} {10}},\ \bibinfo {eid} {065} (\bibinfo {year} {2008})},\
  \Eprint {http://arxiv.org/abs/0808.2096} {arXiv:0808.2096 [hep-th]}
  \BibitemShut {NoStop}%
\bibitem [{\citenamefont {{Susskind}}(2011)}]{add}%
  \BibitemOpen
  \bibfield  {author} {\bibinfo {author} {\bibfnamefont {L.}~\bibnamefont
  {{Susskind}}},\ }\href@noop {} {\bibfield  {journal} {\bibinfo  {journal}
  {ArXiv e-prints}\ } (\bibinfo {year} {2011})},\ \Eprint
  {http://arxiv.org/abs/1101.6048} {arXiv:1101.6048 [hep-th]} \BibitemShut
  {NoStop}%
\bibitem [{\citenamefont {Furuya}\ \emph {et~al.}(1998)\citenamefont {Furuya},
  \citenamefont {Nemes},\ and\ \citenamefont
  {Pellegrino}}]{PhysRevLett.80.5524}%
  \BibitemOpen
  \bibfield  {author} {\bibinfo {author} {\bibfnamefont {K.}~\bibnamefont
  {Furuya}}, \bibinfo {author} {\bibfnamefont {M.~C.}\ \bibnamefont {Nemes}}, \
  and\ \bibinfo {author} {\bibfnamefont {G.~Q.}\ \bibnamefont {Pellegrino}},\
  }\href {\doibase 10.1103/PhysRevLett.80.5524} {\bibfield  {journal} {\bibinfo
   {journal} {Phys. Rev. Lett.}\ }\textbf {\bibinfo {volume} {80}},\ \bibinfo
  {pages} {5524} (\bibinfo {year} {1998})}\BibitemShut {NoStop}%
\bibitem [{\citenamefont {Lakshminarayan}(2001)}]{PhysRevE.64.036207}%
  \BibitemOpen
  \bibfield  {author} {\bibinfo {author} {\bibfnamefont {A.}~\bibnamefont
  {Lakshminarayan}},\ }\href {\doibase 10.1103/PhysRevE.64.036207} {\bibfield
  {journal} {\bibinfo  {journal} {Phys. Rev. E}\ }\textbf {\bibinfo {volume}
  {64}},\ \bibinfo {pages} {036207} (\bibinfo {year} {2001})}\BibitemShut
  {NoStop}%
\bibitem [{\citenamefont {Hosur}\ \emph {et~al.}(2016)\citenamefont {Hosur},
  \citenamefont {Qi}, \citenamefont {Roberts},\ and\ \citenamefont
  {Yoshida}}]{chaos}%
  \BibitemOpen
  \bibfield  {author} {\bibinfo {author} {\bibfnamefont {P.}~\bibnamefont
  {Hosur}}, \bibinfo {author} {\bibfnamefont {X.-L.}\ \bibnamefont {Qi}},
  \bibinfo {author} {\bibfnamefont {D.~A.}\ \bibnamefont {Roberts}}, \ and\
  \bibinfo {author} {\bibfnamefont {B.}~\bibnamefont {Yoshida}},\ }\href
  {\doibase 10.1007/JHEP02(2016)004} {\bibfield  {journal} {\bibinfo  {journal}
  {Journal of High Energy Physics}\ }\textbf {\bibinfo {volume} {2016}},\
  \bibinfo {pages} {1} (\bibinfo {year} {2016})}\BibitemShut {NoStop}%
\bibitem [{\citenamefont {Nandkishore}\ and\ \citenamefont
  {Huse}(2015)}]{mblrev}%
  \BibitemOpen
  \bibfield  {author} {\bibinfo {author} {\bibfnamefont {R.}~\bibnamefont
  {Nandkishore}}\ and\ \bibinfo {author} {\bibfnamefont {D.~A.}\ \bibnamefont
  {Huse}},\ }\href {\doibase 10.1146/annurev-conmatphys-031214-014726}
  {\bibfield  {journal} {\bibinfo  {journal} {Annual Review of Condensed Matter
  Physics}\ }\textbf {\bibinfo {volume} {6}},\ \bibinfo {pages} {15} (\bibinfo
  {year} {2015})}\BibitemShut {NoStop}%
\bibitem [{\citenamefont {Popescu}\ \emph {et~al.}(2006)\citenamefont
  {Popescu}, \citenamefont {Short},\ and\ \citenamefont {Winter}}]{popescu}%
  \BibitemOpen
  \bibfield  {author} {\bibinfo {author} {\bibfnamefont {S.}~\bibnamefont
  {Popescu}}, \bibinfo {author} {\bibfnamefont {A.~J.}\ \bibnamefont {Short}},
  \ and\ \bibinfo {author} {\bibfnamefont {A.}~\bibnamefont {Winter}},\
  }\href@noop {} {\bibfield  {journal} {\bibinfo  {journal} {Nature Physics}\
  }\textbf {\bibinfo {volume} {2}},\ \bibinfo {pages} {754} (\bibinfo {year}
  {2006})}\BibitemShut {NoStop}%
\bibitem [{\citenamefont {Terhal}\ \emph {et~al.}(2001)\citenamefont {Terhal},
  \citenamefont {DiVincenzo},\ and\ \citenamefont
  {Leung}}]{PhysRevLett.86.5807}%
  \BibitemOpen
  \bibfield  {author} {\bibinfo {author} {\bibfnamefont {B.~M.}\ \bibnamefont
  {Terhal}}, \bibinfo {author} {\bibfnamefont {D.~P.}\ \bibnamefont
  {DiVincenzo}}, \ and\ \bibinfo {author} {\bibfnamefont {D.~W.}\ \bibnamefont
  {Leung}},\ }\href {\doibase 10.1103/PhysRevLett.86.5807} {\bibfield
  {journal} {\bibinfo  {journal} {Phys. Rev. Lett.}\ }\textbf {\bibinfo
  {volume} {86}},\ \bibinfo {pages} {5807} (\bibinfo {year}
  {2001})}\BibitemShut {NoStop}%
\bibitem [{\citenamefont {DiVincenzo}\ \emph {et~al.}(2002)\citenamefont
  {DiVincenzo}, \citenamefont {Leung},\ and\ \citenamefont {Terhal}}]{dh02}%
  \BibitemOpen
  \bibfield  {author} {\bibinfo {author} {\bibfnamefont {D.~P.}\ \bibnamefont
  {DiVincenzo}}, \bibinfo {author} {\bibfnamefont {D.~W.}\ \bibnamefont
  {Leung}}, \ and\ \bibinfo {author} {\bibfnamefont {B.~M.}\ \bibnamefont
  {Terhal}},\ }\href {\doibase 10.1109/18.985948} {\bibfield  {journal}
  {\bibinfo  {journal} {IEEE Transactions on Information Theory}\ }\textbf
  {\bibinfo {volume} {48}},\ \bibinfo {pages} {580} (\bibinfo {year}
  {2002})}\BibitemShut {NoStop}%
\bibitem [{\citenamefont {{Lubkin}}(1978)}]{lubkin}%
  \BibitemOpen
  \bibfield  {author} {\bibinfo {author} {\bibfnamefont {E.}~\bibnamefont
  {{Lubkin}}},\ }\href {\doibase 10.1063/1.523763} {\bibfield  {journal}
  {\bibinfo  {journal} {Journal of Mathematical Physics}\ }\textbf {\bibinfo
  {volume} {19}},\ \bibinfo {pages} {1028} (\bibinfo {year}
  {1978})}\BibitemShut {NoStop}%
\bibitem [{\citenamefont {{Lloyd}}\ and\ \citenamefont
  {{Pagels}}(1988)}]{lloydpagels}%
  \BibitemOpen
  \bibfield  {author} {\bibinfo {author} {\bibfnamefont {S.}~\bibnamefont
  {{Lloyd}}}\ and\ \bibinfo {author} {\bibfnamefont {H.}~\bibnamefont
  {{Pagels}}},\ }\href {\doibase 10.1016/0003-4916(88)90094-2} {\bibfield
  {journal} {\bibinfo  {journal} {Annals of Physics}\ }\textbf {\bibinfo
  {volume} {188}},\ \bibinfo {pages} {186} (\bibinfo {year}
  {1988})}\BibitemShut {NoStop}%
\bibitem [{\citenamefont {Foong}\ and\ \citenamefont {Kanno}(1994)}]{FoonK94}%
  \BibitemOpen
  \bibfield  {author} {\bibinfo {author} {\bibfnamefont {S.~K.}\ \bibnamefont
  {Foong}}\ and\ \bibinfo {author} {\bibfnamefont {S.}~\bibnamefont {Kanno}},\
  }\href {\doibase 10.1103/PhysRevLett.72.1148} {\bibfield  {journal} {\bibinfo
   {journal} {Phys. Rev. Lett.}\ }\textbf {\bibinfo {volume} {72}},\ \bibinfo
  {pages} {1148} (\bibinfo {year} {1994})}\BibitemShut {NoStop}%
\bibitem [{\citenamefont {S\'anchez-Ruiz}(1995)}]{Sanc95}%
  \BibitemOpen
  \bibfield  {author} {\bibinfo {author} {\bibfnamefont {J.}~\bibnamefont
  {S\'anchez-Ruiz}},\ }\href {\doibase 10.1103/PhysRevE.52.5653} {\bibfield
  {journal} {\bibinfo  {journal} {Phys. Rev. E}\ }\textbf {\bibinfo {volume}
  {52}},\ \bibinfo {pages} {5653} (\bibinfo {year} {1995})}\BibitemShut
  {NoStop}%
\bibitem [{\citenamefont {Sen}(1996)}]{Sen96}%
  \BibitemOpen
  \bibfield  {author} {\bibinfo {author} {\bibfnamefont {S.}~\bibnamefont
  {Sen}},\ }\href {\doibase 10.1103/PhysRevLett.77.1} {\bibfield  {journal}
  {\bibinfo  {journal} {Phys. Rev. Lett.}\ }\textbf {\bibinfo {volume} {77}},\
  \bibinfo {pages} {1} (\bibinfo {year} {1996})}\BibitemShut {NoStop}%
\bibitem [{\citenamefont {{Knill}}(1995)}]{1995quant.ph..8006K}%
  \BibitemOpen
  \bibfield  {author} {\bibinfo {author} {\bibfnamefont {E.}~\bibnamefont
  {{Knill}}},\ }\href@noop {} {\bibfield  {journal} {\bibinfo  {journal}
  {eprint arXiv:quant-ph/9508006}\ } (\bibinfo {year} {1995})}\BibitemShut
  {NoStop}%
\bibitem [{\citenamefont {Brand\~ao}\ \emph {et~al.}(2016)\citenamefont
  {Brand\~ao}, \citenamefont {Harrow},\ and\ \citenamefont
  {Horodecki}}]{brandaoharrow2}%
  \BibitemOpen
  \bibfield  {author} {\bibinfo {author} {\bibfnamefont {F.~G. S.~L.}\
  \bibnamefont {Brand\~ao}}, \bibinfo {author} {\bibfnamefont {A.~W.}\
  \bibnamefont {Harrow}}, \ and\ \bibinfo {author} {\bibfnamefont
  {M.}~\bibnamefont {Horodecki}},\ }\href {\doibase
  10.1103/PhysRevLett.116.170502} {\bibfield  {journal} {\bibinfo  {journal}
  {Phys. Rev. Lett.}\ }\textbf {\bibinfo {volume} {116}},\ \bibinfo {pages}
  {170502} (\bibinfo {year} {2016})}\BibitemShut {NoStop}%
\bibitem [{\citenamefont {T\'oth}\ and\ \citenamefont
  {Garc\'{\i}a-Ripoll}(2007)}]{PhysRevA.75.042311}%
  \BibitemOpen
  \bibfield  {author} {\bibinfo {author} {\bibfnamefont {G.}~\bibnamefont
  {T\'oth}}\ and\ \bibinfo {author} {\bibfnamefont {J.~J.}\ \bibnamefont
  {Garc\'{\i}a-Ripoll}},\ }\href {\doibase 10.1103/PhysRevA.75.042311}
  {\bibfield  {journal} {\bibinfo  {journal} {Phys. Rev. A}\ }\textbf {\bibinfo
  {volume} {75}},\ \bibinfo {pages} {042311} (\bibinfo {year}
  {2007})}\BibitemShut {NoStop}%
\bibitem [{\citenamefont {Nakata}\ \emph {et~al.}(2017)\citenamefont {Nakata},
  \citenamefont {Hirche}, \citenamefont {Koashi},\ and\ \citenamefont
  {Winter}}]{nakata}%
  \BibitemOpen
  \bibfield  {author} {\bibinfo {author} {\bibfnamefont {Y.}~\bibnamefont
  {Nakata}}, \bibinfo {author} {\bibfnamefont {C.}~\bibnamefont {Hirche}},
  \bibinfo {author} {\bibfnamefont {M.}~\bibnamefont {Koashi}}, \ and\ \bibinfo
  {author} {\bibfnamefont {A.}~\bibnamefont {Winter}},\ }\href {\doibase
  10.1103/PhysRevX.7.021006} {\bibfield  {journal} {\bibinfo  {journal} {Phys.
  Rev. X}\ }\textbf {\bibinfo {volume} {7}},\ \bibinfo {pages} {021006}
  (\bibinfo {year} {2017})}\BibitemShut {NoStop}%
\bibitem [{\citenamefont {Lashkari}\ \emph {et~al.}(2013)\citenamefont
  {Lashkari}, \citenamefont {Stanford}, \citenamefont {Hastings}, \citenamefont
  {Osborne},\ and\ \citenamefont {Hayden}}]{Lashkari2013}%
  \BibitemOpen
  \bibfield  {author} {\bibinfo {author} {\bibfnamefont {N.}~\bibnamefont
  {Lashkari}}, \bibinfo {author} {\bibfnamefont {D.}~\bibnamefont {Stanford}},
  \bibinfo {author} {\bibfnamefont {M.}~\bibnamefont {Hastings}}, \bibinfo
  {author} {\bibfnamefont {T.}~\bibnamefont {Osborne}}, \ and\ \bibinfo
  {author} {\bibfnamefont {P.}~\bibnamefont {Hayden}},\ }\href {\doibase
  10.1007/JHEP04(2013)022} {\bibfield  {journal} {\bibinfo  {journal} {Journal
  of High Energy Physics}\ }\textbf {\bibinfo {volume} {2013}},\ \bibinfo
  {pages} {22} (\bibinfo {year} {2013})}\BibitemShut {NoStop}%
\bibitem [{\citenamefont {Liu}\ \emph {et~al.}(2018)\citenamefont {Liu},
  \citenamefont {Lloyd}, \citenamefont {Zhu},\ and\ \citenamefont
  {Zhu}}]{Liu2018}%
  \BibitemOpen
  \bibfield  {author} {\bibinfo {author} {\bibfnamefont {Z.-W.}\ \bibnamefont
  {Liu}}, \bibinfo {author} {\bibfnamefont {S.}~\bibnamefont {Lloyd}}, \bibinfo
  {author} {\bibfnamefont {E.}~\bibnamefont {Zhu}}, \ and\ \bibinfo {author}
  {\bibfnamefont {H.}~\bibnamefont {Zhu}},\ }\href {\doibase
  10.1007/JHEP07(2018)041} {\bibfield  {journal} {\bibinfo  {journal} {Journal
  of High Energy Physics}\ }\textbf {\bibinfo {volume} {2018}},\ \bibinfo
  {pages} {41} (\bibinfo {year} {2018})}\BibitemShut {NoStop}%
\bibitem [{\citenamefont {Magesan}\ \emph {et~al.}(2011)\citenamefont
  {Magesan}, \citenamefont {Gambetta},\ and\ \citenamefont
  {Emerson}}]{PhysRevLett.106.180504}%
  \BibitemOpen
  \bibfield  {author} {\bibinfo {author} {\bibfnamefont {E.}~\bibnamefont
  {Magesan}}, \bibinfo {author} {\bibfnamefont {J.~M.}\ \bibnamefont
  {Gambetta}}, \ and\ \bibinfo {author} {\bibfnamefont {J.}~\bibnamefont
  {Emerson}},\ }\href {\doibase 10.1103/PhysRevLett.106.180504} {\bibfield
  {journal} {\bibinfo  {journal} {Phys. Rev. Lett.}\ }\textbf {\bibinfo
  {volume} {106}},\ \bibinfo {pages} {180504} (\bibinfo {year}
  {2011})}\BibitemShut {NoStop}%
\bibitem [{\citenamefont {Wallman}\ and\ \citenamefont
  {Flammia}(2014)}]{1367-2630-16-10-103032}%
  \BibitemOpen
  \bibfield  {author} {\bibinfo {author} {\bibfnamefont {J.~J.}\ \bibnamefont
  {Wallman}}\ and\ \bibinfo {author} {\bibfnamefont {S.~T.}\ \bibnamefont
  {Flammia}},\ }\href {http://stacks.iop.org/1367-2630/16/i=10/a=103032}
  {\bibfield  {journal} {\bibinfo  {journal} {New J.\ Phys.}\ }\textbf
  {\bibinfo {volume} {16}},\ \bibinfo {pages} {103032} (\bibinfo {year}
  {2014})}\BibitemShut {NoStop}%
\bibitem [{\citenamefont {Szehr}\ \emph {et~al.}(2013)\citenamefont {Szehr},
  \citenamefont {Dupuis}, \citenamefont {Tomamichel},\ and\ \citenamefont
  {Renner}}]{1367-2630-15-5-053022}%
  \BibitemOpen
  \bibfield  {author} {\bibinfo {author} {\bibfnamefont {O.}~\bibnamefont
  {Szehr}}, \bibinfo {author} {\bibfnamefont {F.}~\bibnamefont {Dupuis}},
  \bibinfo {author} {\bibfnamefont {M.}~\bibnamefont {Tomamichel}}, \ and\
  \bibinfo {author} {\bibfnamefont {R.}~\bibnamefont {Renner}},\ }\href
  {http://stacks.iop.org/1367-2630/15/i=5/a=053022} {\bibfield  {journal}
  {\bibinfo  {journal} {New J.\ Phys.}\ }\textbf {\bibinfo {volume} {15}},\
  \bibinfo {pages} {053022} (\bibinfo {year} {2013})}\BibitemShut {NoStop}%
\bibitem [{\citenamefont {{Zhu}}\ \emph {et~al.}(2016)\citenamefont {{Zhu}},
  \citenamefont {{Kueng}}, \citenamefont {{Grassl}},\ and\ \citenamefont
  {{Gross}}}]{rep}%
  \BibitemOpen
  \bibfield  {author} {\bibinfo {author} {\bibfnamefont {H.}~\bibnamefont
  {{Zhu}}}, \bibinfo {author} {\bibfnamefont {R.}~\bibnamefont {{Kueng}}},
  \bibinfo {author} {\bibfnamefont {M.}~\bibnamefont {{Grassl}}}, \ and\
  \bibinfo {author} {\bibfnamefont {D.}~\bibnamefont {{Gross}}},\ }\href@noop
  {} {\bibfield  {journal} {\bibinfo  {journal} {ArXiv e-prints}\ } (\bibinfo
  {year} {2016})},\ \Eprint {http://arxiv.org/abs/1609.08172} {arXiv:1609.08172
  [quant-ph]} \BibitemShut {NoStop}%
\bibitem [{Note1()}]{Note1}%
  \BibitemOpen
  \bibinfo {note} {Every element of the symmetric group can be uniquely
  decomposed into a product of disjoint cycles (up to relabeling).}\BibitemShut
  {Stop}%
\bibitem [{\citenamefont {{\.{Z}}yczkowski}\ and\ \citenamefont
  {Sommers}(2001)}]{ZyczS01}%
  \BibitemOpen
  \bibfield  {author} {\bibinfo {author} {\bibfnamefont {K.}~\bibnamefont
  {{\.{Z}}yczkowski}}\ and\ \bibinfo {author} {\bibfnamefont {H.-J.}\
  \bibnamefont {Sommers}},\ }\href@noop {} {\bibfield  {journal} {\bibinfo
  {journal} {J. Phys. A: Math. Gen.}\ }\textbf {\bibinfo {volume} {34}},\
  \bibinfo {pages} {7111} (\bibinfo {year} {2001})}\BibitemShut {NoStop}%
\bibitem [{\citenamefont {Malacarne}\ \emph {et~al.}(2002)\citenamefont
  {Malacarne}, \citenamefont {Mendes},\ and\ \citenamefont {Lenzi}}]{MalaML02}%
  \BibitemOpen
  \bibfield  {author} {\bibinfo {author} {\bibfnamefont {L.~C.}\ \bibnamefont
  {Malacarne}}, \bibinfo {author} {\bibfnamefont {R.~S.}\ \bibnamefont
  {Mendes}}, \ and\ \bibinfo {author} {\bibfnamefont {E.~K.}\ \bibnamefont
  {Lenzi}},\ }\href {\doibase 10.1103/PhysRevE.65.046131} {\bibfield  {journal}
  {\bibinfo  {journal} {Phys. Rev. E}\ }\textbf {\bibinfo {volume} {65}},\
  \bibinfo {pages} {046131} (\bibinfo {year} {2002})}\BibitemShut {NoStop}%
\bibitem [{\citenamefont {Collins}\ and\ \citenamefont
  {Nechita}(2010)}]{Collins2010}%
  \BibitemOpen
  \bibfield  {author} {\bibinfo {author} {\bibfnamefont {B.}~\bibnamefont
  {Collins}}\ and\ \bibinfo {author} {\bibfnamefont {I.}~\bibnamefont
  {Nechita}},\ }\href {\doibase 10.1007/s00220-010-1012-0} {\bibfield
  {journal} {\bibinfo  {journal} {Commun. Math. Phys.}\ }\textbf {\bibinfo
  {volume} {297}},\ \bibinfo {pages} {345} (\bibinfo {year}
  {2010})}\BibitemShut {NoStop}%
\bibitem [{\citenamefont {Collins}\ and\ \citenamefont
  {Nechita}(2011)}]{CollN11}%
  \BibitemOpen
  \bibfield  {author} {\bibinfo {author} {\bibfnamefont {B.}~\bibnamefont
  {Collins}}\ and\ \bibinfo {author} {\bibfnamefont {I.}~\bibnamefont
  {Nechita}},\ }\href {\doibase 10.1214/10-AAP722} {\bibfield  {journal}
  {\bibinfo  {journal} {Ann. Appl. Probab.}\ }\textbf {\bibinfo {volume}
  {21}},\ \bibinfo {pages} {1136} (\bibinfo {year} {2011})}\BibitemShut
  {NoStop}%
\bibitem [{\citenamefont {Nica}\ and\ \citenamefont
  {Speicher}(2006)}]{nica2006lectures}%
  \BibitemOpen
  \bibfield  {author} {\bibinfo {author} {\bibfnamefont {A.}~\bibnamefont
  {Nica}}\ and\ \bibinfo {author} {\bibfnamefont {R.}~\bibnamefont
  {Speicher}},\ }\href@noop {} {\emph {\bibinfo {title} {Lectures on the
  Combinatorics of Free Probability}}}\ (\bibinfo  {publisher} {Cambridge
  University Press},\ \bibinfo {year} {2006})\BibitemShut {NoStop}%
\bibitem [{\citenamefont {Ding}\ \emph {et~al.}(2016)\citenamefont {Ding},
  \citenamefont {Hayden},\ and\ \citenamefont {Walter}}]{ding}%
  \BibitemOpen
  \bibfield  {author} {\bibinfo {author} {\bibfnamefont {D.}~\bibnamefont
  {Ding}}, \bibinfo {author} {\bibfnamefont {P.}~\bibnamefont {Hayden}}, \ and\
  \bibinfo {author} {\bibfnamefont {M.}~\bibnamefont {Walter}},\ }\href@noop {}
  {\bibfield  {journal} {\bibinfo  {journal} {Journal of High Energy Physics}\
  }\textbf {\bibinfo {volume} {2016}},\ \bibinfo {pages} {145} (\bibinfo {year}
  {2016})}\BibitemShut {NoStop}%
\bibitem [{\citenamefont {Collins}(2003)}]{collins}%
  \BibitemOpen
  \bibfield  {author} {\bibinfo {author} {\bibfnamefont {B.}~\bibnamefont
  {Collins}},\ }\href {\doibase 10.1155/S107379280320917X} {\bibfield
  {journal} {\bibinfo  {journal} {International Mathematics Research Notices}\
  }\textbf {\bibinfo {volume} {2003}},\ \bibinfo {pages} {953} (\bibinfo {year}
  {2003})}\BibitemShut {NoStop}%
\bibitem [{\citenamefont {Collins}\ and\ \citenamefont
  {Matsumoto}(2009)}]{collins2}%
  \BibitemOpen
  \bibfield  {author} {\bibinfo {author} {\bibfnamefont {B.}~\bibnamefont
  {Collins}}\ and\ \bibinfo {author} {\bibfnamefont {S.}~\bibnamefont
  {Matsumoto}},\ }\href {\doibase 10.1063/1.3251304} {\bibfield  {journal}
  {\bibinfo  {journal} {Journal of Mathematical Physics}\ }\textbf {\bibinfo
  {volume} {50}},\ \bibinfo {pages} {113516} (\bibinfo {year}
  {2009})}\BibitemShut {NoStop}%
\bibitem [{\citenamefont {Zinn-Justin}(2010)}]{zinn}%
  \BibitemOpen
  \bibfield  {author} {\bibinfo {author} {\bibfnamefont {P.}~\bibnamefont
  {Zinn-Justin}},\ }\href {\doibase 10.1007/s11005-009-0365-9} {\bibfield
  {journal} {\bibinfo  {journal} {Letters in Mathematical Physics}\ }\textbf
  {\bibinfo {volume} {91}},\ \bibinfo {pages} {119} (\bibinfo {year}
  {2010})}\BibitemShut {NoStop}%
\bibitem [{\citenamefont {Roberts}\ and\ \citenamefont
  {Yoshida}(2017)}]{Roberts2017}%
  \BibitemOpen
  \bibfield  {author} {\bibinfo {author} {\bibfnamefont {D.~A.}\ \bibnamefont
  {Roberts}}\ and\ \bibinfo {author} {\bibfnamefont {B.}~\bibnamefont
  {Yoshida}},\ }\href {\doibase 10.1007/JHEP04(2017)121} {\bibfield  {journal}
  {\bibinfo  {journal} {Journal of High Energy Physics}\ }\textbf {\bibinfo
  {volume} {2017}},\ \bibinfo {pages} {121} (\bibinfo {year}
  {2017})}\BibitemShut {NoStop}%
\bibitem [{\citenamefont {Shenker}\ and\ \citenamefont
  {Stanford}(2014{\natexlab{a}})}]{Shenker2014}%
  \BibitemOpen
  \bibfield  {author} {\bibinfo {author} {\bibfnamefont {S.~H.}\ \bibnamefont
  {Shenker}}\ and\ \bibinfo {author} {\bibfnamefont {D.}~\bibnamefont
  {Stanford}},\ }\href {\doibase 10.1007/JHEP03(2014)067} {\bibfield  {journal}
  {\bibinfo  {journal} {Journal of High Energy Physics}\ }\textbf {\bibinfo
  {volume} {2014}},\ \bibinfo {pages} {67} (\bibinfo {year}
  {2014}{\natexlab{a}})}\BibitemShut {NoStop}%
\bibitem [{\citenamefont {Shenker}\ and\ \citenamefont
  {Stanford}(2014{\natexlab{b}})}]{Shenker20142}%
  \BibitemOpen
  \bibfield  {author} {\bibinfo {author} {\bibfnamefont {S.~H.}\ \bibnamefont
  {Shenker}}\ and\ \bibinfo {author} {\bibfnamefont {D.}~\bibnamefont
  {Stanford}},\ }\href {\doibase 10.1007/JHEP12(2014)046} {\bibfield  {journal}
  {\bibinfo  {journal} {Journal of High Energy Physics}\ }\textbf {\bibinfo
  {volume} {2014}},\ \bibinfo {pages} {46} (\bibinfo {year}
  {2014}{\natexlab{b}})}\BibitemShut {NoStop}%
\bibitem [{\citenamefont {Shenker}\ and\ \citenamefont
  {Stanford}(2015)}]{Shenker2015}%
  \BibitemOpen
  \bibfield  {author} {\bibinfo {author} {\bibfnamefont {S.~H.}\ \bibnamefont
  {Shenker}}\ and\ \bibinfo {author} {\bibfnamefont {D.}~\bibnamefont
  {Stanford}},\ }\href {\doibase 10.1007/JHEP05(2015)132} {\bibfield  {journal}
  {\bibinfo  {journal} {Journal of High Energy Physics}\ }\textbf {\bibinfo
  {volume} {2015}},\ \bibinfo {pages} {132} (\bibinfo {year}
  {2015})}\BibitemShut {NoStop}%
\bibitem [{\citenamefont {Roberts}\ \emph {et~al.}(2015)\citenamefont
  {Roberts}, \citenamefont {Stanford},\ and\ \citenamefont
  {Susskind}}]{Roberts2015}%
  \BibitemOpen
  \bibfield  {author} {\bibinfo {author} {\bibfnamefont {D.~A.}\ \bibnamefont
  {Roberts}}, \bibinfo {author} {\bibfnamefont {D.}~\bibnamefont {Stanford}}, \
  and\ \bibinfo {author} {\bibfnamefont {L.}~\bibnamefont {Susskind}},\ }\href
  {\doibase 10.1007/JHEP03(2015)051} {\bibfield  {journal} {\bibinfo  {journal}
  {Journal of High Energy Physics}\ }\textbf {\bibinfo {volume} {2015}},\
  \bibinfo {pages} {51} (\bibinfo {year} {2015})}\BibitemShut {NoStop}%
\bibitem [{\citenamefont {Roberts}\ and\ \citenamefont
  {Stanford}(2015)}]{PhysRevLett.115.131603}%
  \BibitemOpen
  \bibfield  {author} {\bibinfo {author} {\bibfnamefont {D.~A.}\ \bibnamefont
  {Roberts}}\ and\ \bibinfo {author} {\bibfnamefont {D.}~\bibnamefont
  {Stanford}},\ }\href {\doibase 10.1103/PhysRevLett.115.131603} {\bibfield
  {journal} {\bibinfo  {journal} {Phys. Rev. Lett.}\ }\textbf {\bibinfo
  {volume} {115}},\ \bibinfo {pages} {131603} (\bibinfo {year}
  {2015})}\BibitemShut {NoStop}%
\bibitem [{\citenamefont {Maldacena}\ \emph {et~al.}(2016)\citenamefont
  {Maldacena}, \citenamefont {Shenker},\ and\ \citenamefont
  {Stanford}}]{Maldacena2016}%
  \BibitemOpen
  \bibfield  {author} {\bibinfo {author} {\bibfnamefont {J.}~\bibnamefont
  {Maldacena}}, \bibinfo {author} {\bibfnamefont {S.~H.}\ \bibnamefont
  {Shenker}}, \ and\ \bibinfo {author} {\bibfnamefont {D.}~\bibnamefont
  {Stanford}},\ }\href {\doibase 10.1007/JHEP08(2016)106} {\bibfield  {journal}
  {\bibinfo  {journal} {Journal of High Energy Physics}\ }\textbf {\bibinfo
  {volume} {2016}},\ \bibinfo {pages} {106} (\bibinfo {year}
  {2016})}\BibitemShut {NoStop}%
\bibitem [{\citenamefont {{Kukuljan}}\ \emph {et~al.}(2017)\citenamefont
  {{Kukuljan}}, \citenamefont {{Grozdanov}},\ and\ \citenamefont
  {{Prosen}}}]{weakchaos}%
  \BibitemOpen
  \bibfield  {author} {\bibinfo {author} {\bibfnamefont {I.}~\bibnamefont
  {{Kukuljan}}}, \bibinfo {author} {\bibfnamefont {S.}~\bibnamefont
  {{Grozdanov}}}, \ and\ \bibinfo {author} {\bibfnamefont {T.}~\bibnamefont
  {{Prosen}}},\ }\href@noop {} {\bibfield  {journal} {\bibinfo  {journal}
  {ArXiv e-prints}\ } (\bibinfo {year} {2017})},\ \Eprint
  {http://arxiv.org/abs/1701.09147} {arXiv:1701.09147 [cond-mat.stat-mech]}
  \BibitemShut {NoStop}%
\bibitem [{Note2()}]{Note2}%
  \BibitemOpen
  \bibinfo {note} {Here ``local'' should mean all interaction terms involve at
  most $k$ qubits where $k$ is some constant ($k$-local), rather than spatially
  local on a finite-dimensional graph. This was not made very clear in \cite
  {nakata}.}\BibitemShut {Stop}%
\bibitem [{\citenamefont {Liu}\ and\ \citenamefont
  {Suh}(2014{\natexlab{a}})}]{tsunami}%
  \BibitemOpen
  \bibfield  {author} {\bibinfo {author} {\bibfnamefont {H.}~\bibnamefont
  {Liu}}\ and\ \bibinfo {author} {\bibfnamefont {S.~J.}\ \bibnamefont {Suh}},\
  }\href {\doibase 10.1103/PhysRevLett.112.011601} {\bibfield  {journal}
  {\bibinfo  {journal} {Phys. Rev. Lett.}\ }\textbf {\bibinfo {volume} {112}},\
  \bibinfo {pages} {011601} (\bibinfo {year} {2014}{\natexlab{a}})}\BibitemShut
  {NoStop}%
\bibitem [{\citenamefont {Liu}\ and\ \citenamefont
  {Suh}(2014{\natexlab{b}})}]{PhysRevD.89.066012}%
  \BibitemOpen
  \bibfield  {author} {\bibinfo {author} {\bibfnamefont {H.}~\bibnamefont
  {Liu}}\ and\ \bibinfo {author} {\bibfnamefont {S.~J.}\ \bibnamefont {Suh}},\
  }\href {\doibase 10.1103/PhysRevD.89.066012} {\bibfield  {journal} {\bibinfo
  {journal} {Phys. Rev. D}\ }\textbf {\bibinfo {volume} {89}},\ \bibinfo
  {pages} {066012} (\bibinfo {year} {2014}{\natexlab{b}})}\BibitemShut
  {NoStop}%
\bibitem [{\citenamefont {Gu}\ \emph {et~al.}(2017)\citenamefont {Gu},
  \citenamefont {Lucas},\ and\ \citenamefont {Qi}}]{Gu2017}%
  \BibitemOpen
  \bibfield  {author} {\bibinfo {author} {\bibfnamefont {Y.}~\bibnamefont
  {Gu}}, \bibinfo {author} {\bibfnamefont {A.}~\bibnamefont {Lucas}}, \ and\
  \bibinfo {author} {\bibfnamefont {X.-L.}\ \bibnamefont {Qi}},\ }\href
  {\doibase 10.1007/JHEP09(2017)120} {\bibfield  {journal} {\bibinfo  {journal}
  {Journal of High Energy Physics}\ }\textbf {\bibinfo {volume} {2017}},\
  \bibinfo {pages} {120} (\bibinfo {year} {2017})}\BibitemShut {NoStop}%
\bibitem [{\citenamefont {Sachdev}\ and\ \citenamefont
  {Ye}(1993)}]{PhysRevLett.70.3339}%
  \BibitemOpen
  \bibfield  {author} {\bibinfo {author} {\bibfnamefont {S.}~\bibnamefont
  {Sachdev}}\ and\ \bibinfo {author} {\bibfnamefont {J.}~\bibnamefont {Ye}},\
  }\href {\doibase 10.1103/PhysRevLett.70.3339} {\bibfield  {journal} {\bibinfo
   {journal} {Phys. Rev. Lett.}\ }\textbf {\bibinfo {volume} {70}},\ \bibinfo
  {pages} {3339} (\bibinfo {year} {1993})}\BibitemShut {NoStop}%
\bibitem [{\citenamefont {Kitaev}(2015)}]{k}%
  \BibitemOpen
  \bibfield  {author} {\bibinfo {author} {\bibfnamefont {A.}~\bibnamefont
  {Kitaev}},\ }\href {http://online.kitp.ucsb.edu/online/entangled15/kitaev/,
  http://online.kitp.ucsb.edu/online/entangled15/kitaev2/.} {} (\bibinfo {year}
  {2015}),\ \bibinfo {note}
  {http://online.kitp.ucsb.edu/online/entangled15/kitaev/,
  http://online.kitp.ucsb.edu/online/entangled15/kitaev2/.}\BibitemShut {Stop}%
\bibitem [{\citenamefont {Brown}\ \emph
  {et~al.}(2016{\natexlab{a}})\citenamefont {Brown}, \citenamefont {Roberts},
  \citenamefont {Susskind}, \citenamefont {Swingle},\ and\ \citenamefont
  {Zhao}}]{cealong}%
  \BibitemOpen
  \bibfield  {author} {\bibinfo {author} {\bibfnamefont {A.~R.}\ \bibnamefont
  {Brown}}, \bibinfo {author} {\bibfnamefont {D.~A.}\ \bibnamefont {Roberts}},
  \bibinfo {author} {\bibfnamefont {L.}~\bibnamefont {Susskind}}, \bibinfo
  {author} {\bibfnamefont {B.}~\bibnamefont {Swingle}}, \ and\ \bibinfo
  {author} {\bibfnamefont {Y.}~\bibnamefont {Zhao}},\ }\href {\doibase
  10.1103/PhysRevD.93.086006} {\bibfield  {journal} {\bibinfo  {journal} {Phys.
  Rev. D}\ }\textbf {\bibinfo {volume} {93}},\ \bibinfo {pages} {086006}
  (\bibinfo {year} {2016}{\natexlab{a}})}\BibitemShut {NoStop}%
\bibitem [{\citenamefont {Brown}\ \emph
  {et~al.}(2016{\natexlab{b}})\citenamefont {Brown}, \citenamefont {Roberts},
  \citenamefont {Susskind}, \citenamefont {Swingle},\ and\ \citenamefont
  {Zhao}}]{CEA}%
  \BibitemOpen
  \bibfield  {author} {\bibinfo {author} {\bibfnamefont {A.~R.}\ \bibnamefont
  {Brown}}, \bibinfo {author} {\bibfnamefont {D.~A.}\ \bibnamefont {Roberts}},
  \bibinfo {author} {\bibfnamefont {L.}~\bibnamefont {Susskind}}, \bibinfo
  {author} {\bibfnamefont {B.}~\bibnamefont {Swingle}}, \ and\ \bibinfo
  {author} {\bibfnamefont {Y.}~\bibnamefont {Zhao}},\ }\href {\doibase
  10.1103/PhysRevLett.116.191301} {\bibfield  {journal} {\bibinfo  {journal}
  {Phys. Rev. Lett.}\ }\textbf {\bibinfo {volume} {116}},\ \bibinfo {pages}
  {191301} (\bibinfo {year} {2016}{\natexlab{b}})}\BibitemShut {NoStop}%
\bibitem [{\citenamefont {{Brown}}\ and\ \citenamefont
  {{Susskind}}(2017)}]{secondlaw}%
  \BibitemOpen
  \bibfield  {author} {\bibinfo {author} {\bibfnamefont {A.~R.}\ \bibnamefont
  {{Brown}}}\ and\ \bibinfo {author} {\bibfnamefont {L.}~\bibnamefont
  {{Susskind}}},\ }\href@noop {} {\bibfield  {journal} {\bibinfo  {journal}
  {ArXiv e-prints}\ } (\bibinfo {year} {2017})},\ \Eprint
  {http://arxiv.org/abs/1701.01107} {arXiv:1701.01107 [hep-th]} \BibitemShut
  {NoStop}%
\end{thebibliography}

%

\end{document}